\documentclass[12pt]{iopart}\usepackage{graphicx,amssymb}
\begin{document}

\title[Geodesic Flows in a Charged Black Hole Spacetime with Quintessence]{Geodesic Flows in a Charged Black Hole Spacetime with Quintessence}

\author{Hemwati Nandan}
\address{Department of Physics, Gurukul Kangri Vishwavidyalaya,
Haridwar, Uttarakhand 249 404, India}
\ead{hnandan@iucaa.ernet.in}

\author{Rashmi Uniyal}
\address{Department of Physics, Government Degree College, Narendranagar, Tehri Garhwal, Uttarakhand 249 175, India}
\ead{rashmiuniyal001@gmail.com}
\address{Department of Physics, Gurukul Kangri Vishwavidyalaya,
Haridwar, Uttarakhand 249 404, India}

\begin{abstract}
\noindent We investigate the evolution of timelike geodesic congruences, in the background of a charged black hole spacetime surrounded with quintessence.
The Raychaudhuri equations for three kinematical quantities namely the expansion scalar, shear and rotation along the geodesic flows in such spacetime are obtained and solved numerically.
We have also analysed both the weak and the strong energy conditions for the focusing of timelike geodesic congruences.
The effect of the normalisation constant ($\alpha$) and equation of state parameter ($\epsilon$) on the evolution of the expansion scalar is discussed, for the congruences with and without an initial shear and rotation.
It is observed that there always exists a critical value of the initial expansion below which we have focusing with smaller values of the normalisation constant and equation of state parameter.
As the corresponding values of both of these parameters are increased, no geodesic focusing is observed.
The results obtained are then compared with those of the Reissener Nordtr$\ddot{o}$m and Schwarzschild  black hole spacetimes as well as their de-Sitter black hole analouges accordingly.
\end{abstract}
\maketitle
\section{Introduction:}
\noindent
The Black Holes (BHs) are the most fascinating objects in the Universe those arise in Einstein's General Relativity (GR), a classical theory of gravity proposed by Einstein just a century ago \cite{Hartle2003,Chandra1992,Poi2004,Wald1984,Schutz1985}.
The Schwarzschild metric obtained in GR was the first unique solution to Einstein’s field equations in vacuum with a spherically symmetric matter distribution, which represents the simplest spacetime of a black hole having mass but no charge and spin \cite{Eins15,Schwar1916}.
There are however other BH spacetimes emerging as solutions of Einstein’s field equations in GR  having charge/or spin with mass such as Reissener-Nordstr$\ddot{o}$m BH \cite{Reissner1916,Nords1918}, Kerr BH \cite{Kerr1963}, Kerr-Newmann BH spacetimes \cite{Newman1965} alongwith the BH spacetimes in other alternative theories of gravity like string theory.\\
\noindent
In GR, the curvature plays crucial role to understand the geometric effect of curved spacetime. The study of geodesics alongwith their deformations in the background of a given spacetime is an elegant way to describe the underlying geometry of that particular spacetime
\cite{Misner1973,Hartle2003,Chandra1992,Poi2004,Wald1984,Schutz1985,Padmanabhan2003}.
A number of studies related to the geodesic motion in the background of various black hole spacetimes have been performed time and again mainly in view of their astrophysical importance \cite{Hack2010,Hec2008,Heck2008,Jamil:2014rsa,Diemer:2014lba,Grunau:2013oca,Fernando:2012ue,Synge:1934zza,Pirani:1956tn,Ellis:1997ey,Ghosh:2009ig,Koley:2003tp,
Uniyal:2014oaa,Uniyal:2014paa,Kuniyal2015,Uniyal:2015sta}.\\
\noindent The observations from supernovae (Type Ia), cosmic microwave background radiation (CMBR), Baryon acoustic oscillations (BAO) and the Hubble measurements indicates that our universe  appears to be expanding at an increasing rate.
The driving force behind such an accelerating universe is believed to be some unknown form of energy with a large negative pressure which is known as dark energy.
There are several candidates for dark energy, such as cosmological constant \cite{Padmanabhan2003,Cunha:2003vg}, phantom \cite{Caldwell:1999ew,Chimento:2003qy,Caldwell:1997ii,Sahni:1999qe}, quintessence \cite{Capozziello:2005ra,Vikman:2004dc}, K-essence \cite{Chiba:1999ka,Scherrer:2004au}
and quintom \cite{Wei:2005nw,Zhao:2005vj,Chimento:2008ws} with various models subjected to the different values of the equation of state (EOS) parameter ($\epsilon$) which relates the energy density to the pressure.
The quintessence scalar field model as an alternative to dark energy is one of the most popular models with the EOS parameter lying in the range $-1<\epsilon<-1/3$.
\noindent It would therefore be quite interesting to study the geodesics and their deformations in the background of a charged BH spacetime surrounded by quintessence to see the effect of dark energy, if any, on the universe locally.
It is also important to look on the matter distribution which causes this spacetime such that the Einstein equations hold and to identify the interesting regions in the spacetime in view of the weak energy condition (WEC) and strong energy condition (SEC).
\\
\noindent In the present paper, we study the geodesic flows and deformations alongwith energy conditions around a charged BH spacetime surrounded by quintessence by solving the evolution (i.e. Raychaudhuri) equations as an \textit{initial value problem} for expansion scalar, shear and rotation (ESR variables) numerically.
First, we briefly review the spacetime used in the next section.
In section III, the nature of effective potential is discussed.
Section IV deals with the discussion of the kinematics of geodesic flows and visualization of ESR.
Finally, the results are summarized in Section V.
\section{The charged black hole spacetime surrounded with\\quintessence}

\noindent We consider a charged BH surrounded by quintessence with the EOS parameter $\epsilon = \frac{p_{\Phi}}{\rho_{\Phi}}$.
For the static spherically-symmetric quintessence surrounding a BH, the energy density of quintessence scalar field ($\Phi$) reduces to the following form \cite{Kiselev:2002dx}
\begin{equation}
\rho_{\Phi} = - \frac{\alpha}{2}\frac{3\epsilon}{r^{3(1+\epsilon)}},
\end{equation}
where the allowed values for $\epsilon$ lies between $-1<\epsilon<-\frac{1}{3}$ \cite{Kiselev:2002dx} and $\alpha$ is the normalization constant. The energy density of scalar field, $\rho_{\Phi}$ is always a positive quantity so $\epsilon$ has a negative value while the normalization factor $\alpha$ should be a positive quantity.
Based on such standpoints, the metric of a charged BH then reads as,
\begin{equation}
ds^2 = f(r)dt^2-\frac{1}{f(r)}dr^2 -r^2(d\theta^2 + sin^2 \theta d\phi^2),
\label{lineelement}
\end{equation}
where $$f(r) = 1-\frac{2M}{r}+\frac{Q^2}{r^2}-\frac{\alpha}{r^{3\epsilon+1}},$$ here $M$ and $Q$ represent the mass and charge of the BH respectively.
The above metric (\ref{lineelement}) represents a BH for $M>Q$, an extremal BH for $M=Q$ and a naked singularity for $M<Q$ ( for a complete discussion on the horizon structure for this spacetime see \cite{Fernando:2014wma}).
The metric reduces to Reissener Nordstr$\ddot{o}$m black hole (RNBH) in the limit $\alpha=0$, which further reduces to Schwarzschild black hole (SBH) in the absence of charge. 
In addition to this, with $\epsilon = -1$, it also reproduces the corresponding BH spacetimes with cosmological constant.
The geodesic equations for the metric (\ref{lineelement}) are given by,

\begin{equation}
\ddot t  + \frac{ f'(r)} {f(r)}\, \dot r \, \dot t  = 0,
\label{tg11}
\end{equation}
\begin{equation}
\ddot r  + \left [\frac{ f'(r) \, \dot t^2 +  f^{-1}(r)' \,\dot r^2  - 2 r \, \dot \theta^2 -2 r \, \sin^2 \theta \, \dot \phi^2} {2 \, f^{-1} (r)} \right] = 0,
\label{rg11}
\end{equation}
\begin{equation}
\ddot \theta + \frac{2}{r}\, \dot r \, \dot \theta - \cos \theta \, \sin \theta\,  \dot \phi^2  = 0,
\label{ps1}
\end{equation}
\begin{equation}
\ddot \phi + \frac{2}{r} \, \dot r \, \dot \phi + 2 \, \cot \theta \, \dot \theta \,  \dot \phi =0,
\label{ph1}
\end{equation}
where the prime denotes the differentiation with respect to $r$.
The first integral of the geodesic Eqs. (\ref{tg11}) and (\ref{ph1}) on equatorial plane (i.e. $\theta= \pi/2$) read as,
\begin{equation}
\dot{t} = \frac{E}{1-\frac{2M}{r}+\frac{Q^2}{r^2}-\frac{\alpha}{r^{3\epsilon+1}}},
\label{eq:tg2}
\end{equation}
\begin{equation}
\dot{\phi} = \frac{L}{r^2},
\label{eq:ph2}
\end{equation}
where $E$ and  $L$ are the integrating constants which correspond to the conserved total energy and angular momentum per unit mass respectively for a test particle.
Using the constraint $u^{\mu}u_{\mu}=-1$, the expression for radial velocity ($u^r$ = $\dot{r}$) can now be obtained as,
\begin{equation}
{\dot{r}}^2={E}^2-V_{eff},
\label{eq:velocity_r}
\end{equation}
where $V_{eff}$ is defined as an effective potential and is expressed as,
\begin{equation}
V_{eff}=\left(1-\frac{2M}{r}+\frac{Q^2}{r^2}-\frac{\alpha}{r^{3\epsilon+1}}\right)\left(\frac{{L}^2}{r^2}+1\right).
\label{eq:Effective-Potential}
\end{equation}
\section{Nature of Effective Potential}

\begin{figure}[h!]
\centerline{
\includegraphics[scale=0.28]{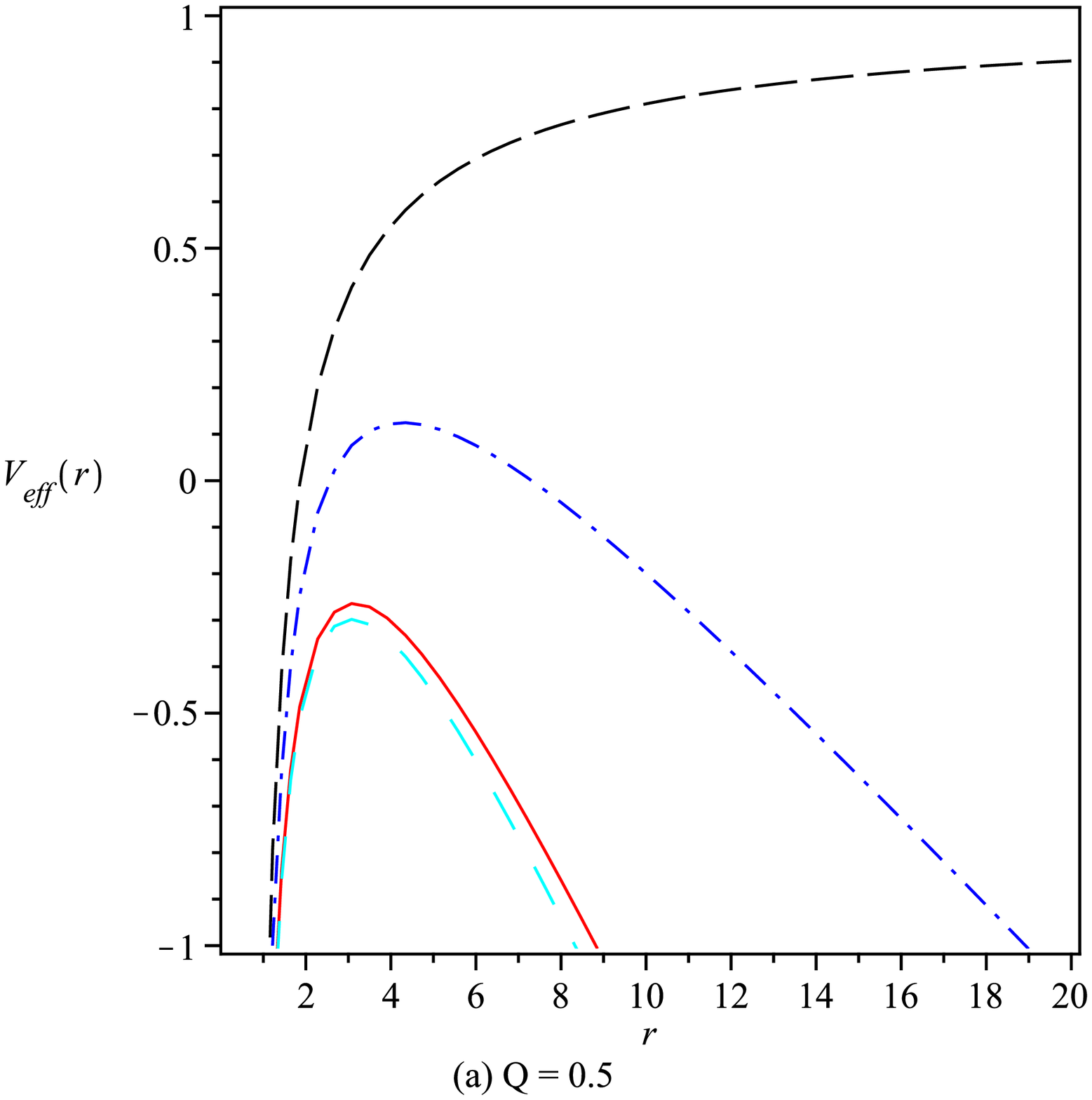}
\includegraphics[scale=0.28]{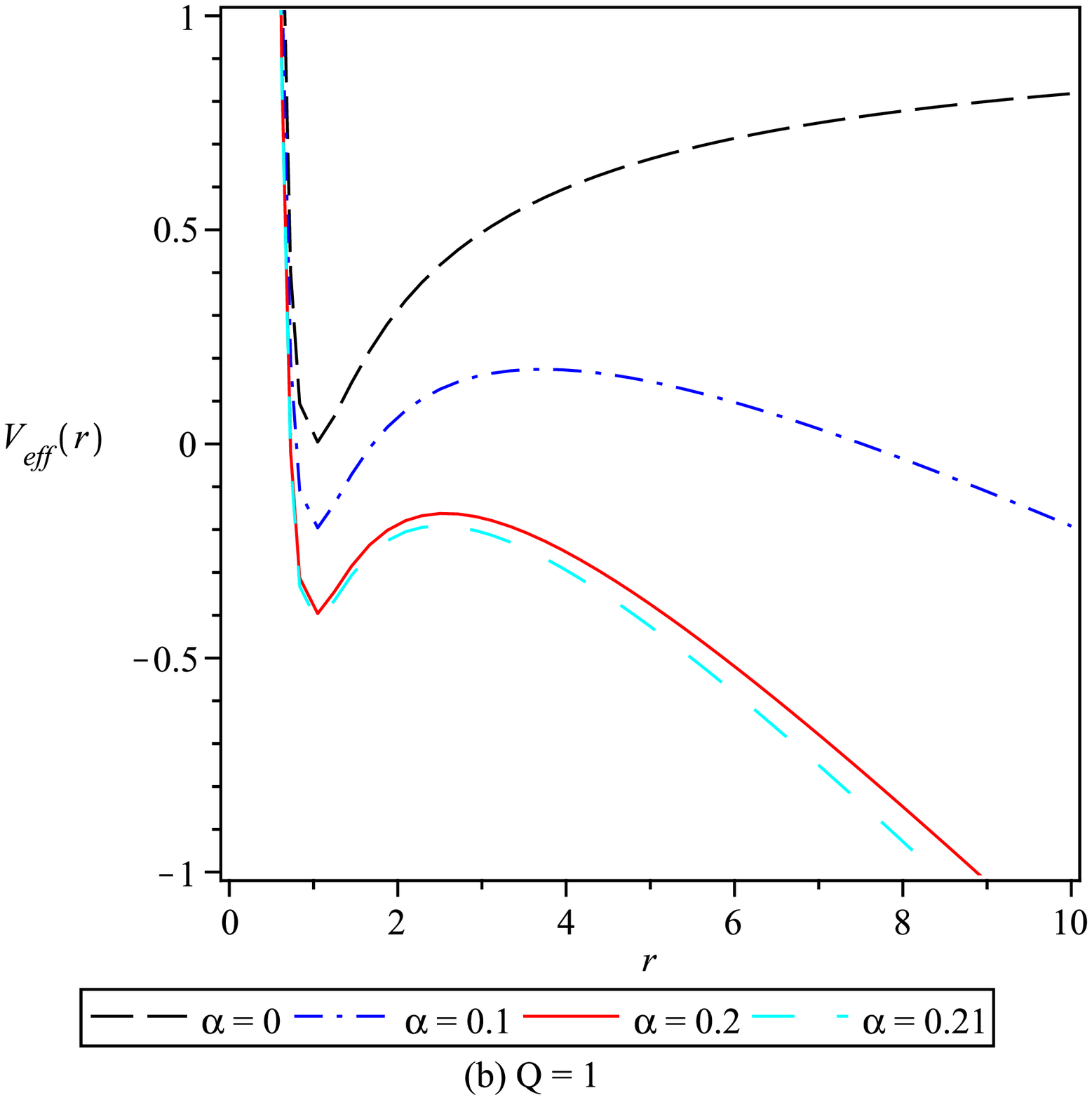}
\includegraphics[scale=0.28]{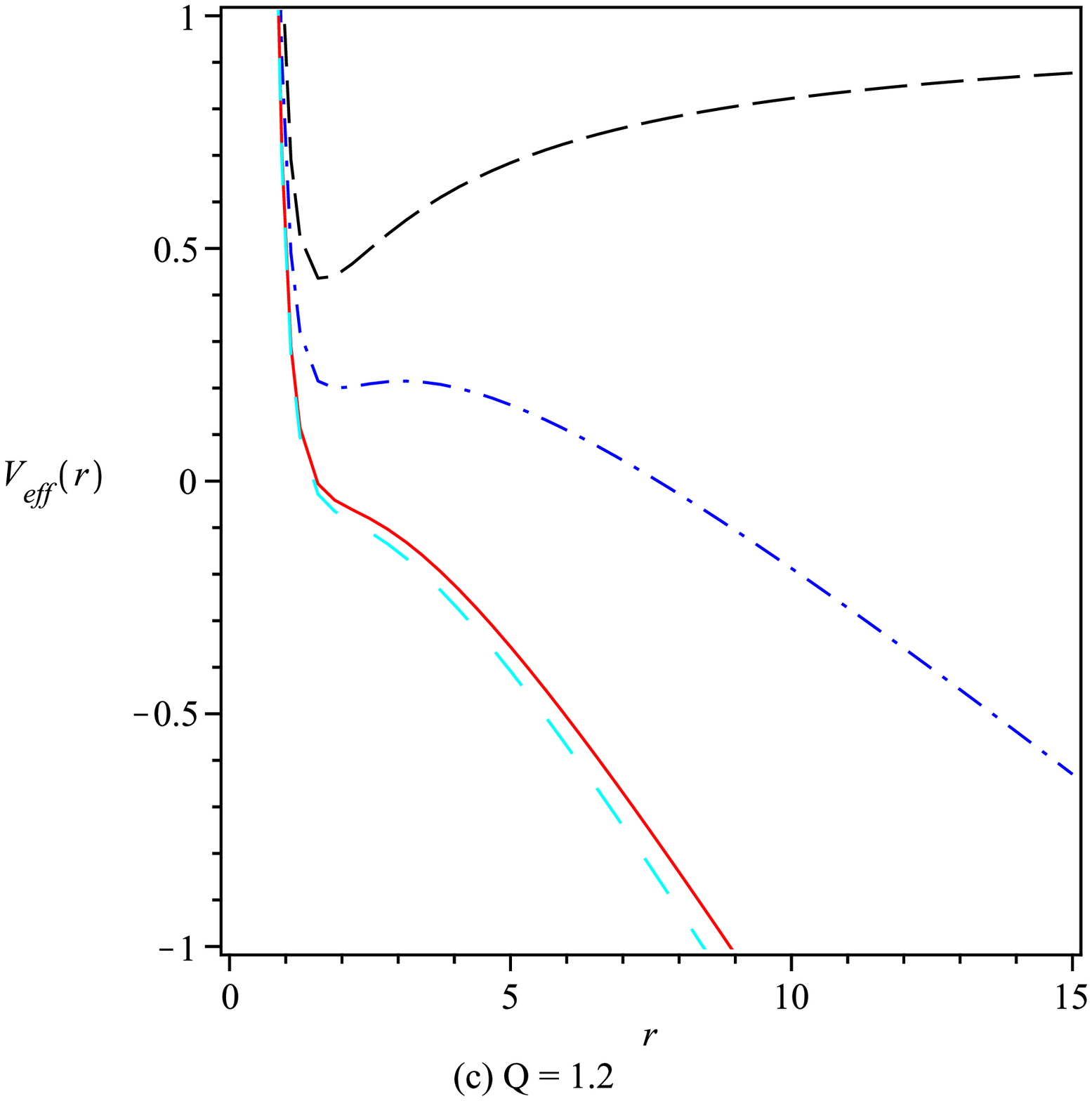}}
\label{fig:Potential-Combo}
\caption{The Effective Potential for $M=L=1$ and $\epsilon=-2/3$.}
\end{figure}
\begin{figure}[h!]
\centerline{
\includegraphics[scale=0.4]{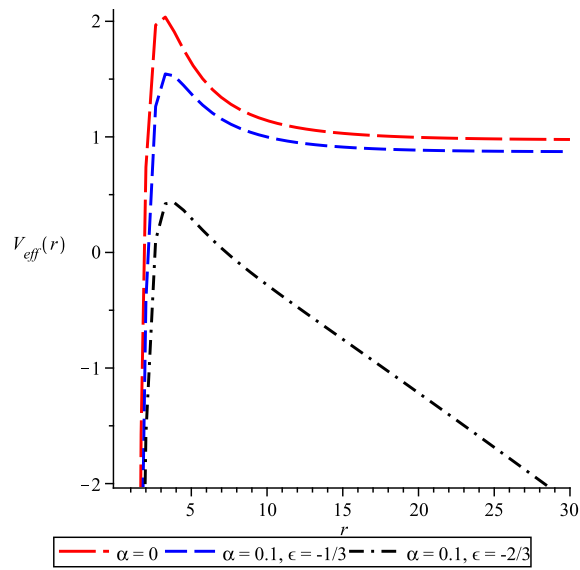}
}
\label{fig:Potential}
\caption{The Effective Potential with $M$ = 1, $Q$ = 0.5, $L$ = 6.5.}
\end{figure}
\noindent Fig.(2) represents the nature of effective potential for different values of EOS parameter.
It is evident from the figure that the nature of effective potential is qualitatively similar for SBH and that of charged BH with $\epsilon=-1/3$. While for $\epsilon=-2/3$, there exists no minima in the radial plot of effective potential.
Hence there are no stable circular orbits for charged BH surrounded with quintessence.
\noindent For the circular motion of test particle, its radial velocity vanishes.
Hence using Eq.(\ref{eq:velocity_r}) and Eq.(\ref{eq:Effective-Potential}) the angular momentum and energy per unit mass for the incoming test particle in circular motion is,
\begin{equation}
PL_c^2 = -r^2_c\left(2M{r^{3\epsilon+2}_c}-2Q^2{r^{3\epsilon+1}_c}+3\alpha\epsilon{r^2_c}+\alpha{r^2_c}\right),
\label{eq:critical-ang-mom}
\end{equation}
\begin{eqnarray}
{P}\,r^2_c E_c^2 = 4Q^2\alpha r^2_c -8M\alpha r^3_c+4\alpha r^4_c-2{r^{3\epsilon+5}_c}-2Q^4 {r^{3\epsilon+1}_c}-4Q^2 {r^{3\epsilon+3}_c} \nonumber\\
\hspace{1.8cm}-2{\alpha}^2 {r^{-3\epsilon+3}_c}+8M{r^{3\epsilon+4}_c}-8M^2{r^{3\epsilon+3}_c}+8Q^2M{r^{3\epsilon+2}_c},
\label{eq:critical-energy}
\end{eqnarray}
where $P = 3\alpha(\epsilon+1)r^2_c -4Q^2 {r^{3\epsilon+1}_c} +6M{r^{3\epsilon+2}_c}-2{r^{3\epsilon+3}_c}$ and $r_c$ represents the radius of circular orbit.
\section{Geodesic Flows}
\subsection{The energy conditions}
\noindent The Ricci scalar for the black hole spacetime given by eq.(\ref{lineelement}) is calculated as,
\begin{equation}
R = \frac{3\alpha\epsilon\left(1-3\epsilon\right)}{r^{3\epsilon+3}},
\label{RicciScalar}
\end{equation}
one may notice that it diverges at $r\rightarrow0$ and vanishes at $r\rightarrow\infty$.
The stress-energy tensor is proportional to,
\begin{eqnarray}
T^{\mu\nu}= -\frac{A}{B{r^2}} \hspace{1cm} 0 \hspace{1cm} 0 \hspace{1cm} 0 \nonumber\\
    \hspace{1.4cm}0 \hspace{1cm} \frac{AB}{r^{6\epsilon+4}} \hspace{1cm} 0 \hspace{1cm} 0 \nonumber\\
    \hspace{1.4cm} 0 \hspace{1.6cm} 0 \hspace{1cm} C \hspace{1cm} 0 \nonumber\\
   \hspace{1.4cm} 0 \hspace{1.6cm} 0 \hspace{1cm} 0 \hspace{1cm} C
\end{eqnarray}

\noindent where,
\vspace{-5mm}
\begin{equation}
A = 3\alpha\epsilon-{Q^2}r^{3\epsilon-1},\, B = r^{3\epsilon+1}-2Mr^{3\epsilon}+{Q^2}r^{3\epsilon-1}-\alpha, \, C = -\frac{3\alpha\epsilon(1+3\epsilon)}{2r^{3\epsilon+5}}.
\end{equation}
In order to analyse the energy conditions, it is convenient to introduce an orthonormal frame that satisfies,
\begin{equation}
g_{\mu\nu}e_{\alpha}^{\mu}e_{\beta}^{\nu}=\eta_{\alpha\beta},
\end{equation}
where $\eta_{\alpha\beta}$ = diag(1,-1,-1,-1) is the Lorentzian metric.
We consider a choice of orthonormal basis for given metric as $e_{\alpha}^{\mu}$ = diag $\left(\frac{1}{\sqrt{g_{00}}},\frac{1}{\sqrt{-g_{11}}},\frac{1}{\sqrt{-g_{22}}},\frac{1}{\sqrt{-g_{33}}}\right)$, where the energy momentum tensor can be written in the following form,
\begin{equation}
T^{\mu\nu}=\rho e_{0}^{\mu}e_{0}^{\nu} + {p_1}e_{1}^{\mu}e_{1}^{\nu} + {p_2}e_{2}^{\mu}e_{2}^{\nu} + {p_3}e_{3}^{\mu}e_{3}^{\nu}.
\end{equation}
The condition for WEC for quintessential fields is,
\begin{equation}
T_{\mu\nu}u^{\mu}u^{\nu}\geq0\approx\rho\geq0, \rho+p_{i}>0,
\label{WEC}
\end{equation}
and for the given spacetime (\ref{lineelement}), we have,
\begin{equation}
T_{\mu\nu}u^{\mu}u^{\nu}=\frac{1}{r^{3\epsilon+3}}\left(Q^2 r^{3\epsilon-1}-3\alpha\epsilon\right).
\end{equation}
Hence the WEC simplifies as follows,
\begin{equation}
Q^2 r^{3\epsilon-1}-3\alpha\epsilon\geq0.
\label{eq:WEC-spt}
\end{equation}
However the SEC reads as,
\begin{equation}
T_{\mu\nu}u^{\mu}u^{\nu}\geq\frac{1}{2}T_\nu^\mu u^{\nu}u_{\nu},
\label{SEC}
\end{equation}
for the spacetime used it can be written as,
\begin{equation}
\frac{1}{2r^{3\epsilon+1}}\left[2Q^2 r^{3\epsilon-1}-3\alpha\epsilon(3\epsilon+1)\right]\geq0,
\label{eq:SEC-spt}
\end{equation}
If WEC presented by Eq.(\ref{eq:WEC-spt}) follows, the expression (\ref{eq:SEC-spt}) for SEC reduces to the following,
\begin{equation}\label{eq:SEC-smp}
  3\alpha\epsilon\left(1-3\epsilon\right)\geq0,
\end{equation}
hence for $\alpha>0$ and $-\frac{1}{3}<\epsilon<-1$, the above condition is clearly violated.
Both the WEC and SEC therefore ensure that $r\neq0$.
Hence, locally the attractive nature of gravity may exist there, but on average the quintessential fields have a repulsive nature, which can be visualised in the deformation of geodesic congruences as discussed below.
\subsection{Raychaudhuri equations for ESR Variables}
\noindent
The spacetime given by Eq.(\ref{lineelement}) can be decomposed into a transverse part i.e. a transverse metric $h_{\mu\nu}$ on a spacelike hypersurface and a longitudinal part ($-u_{\mu}u_{\nu}$) as follows,
\begin{equation}
h_{\mu\nu}=g_{\mu\nu}+u_{\mu}u_{\nu},\,\hspace{10mm} (\mu,\nu=0,1,2,3).
\label{eq:trans_h}
\end{equation}
Here $u^{\mu}$ (a timelike vector field) satisfies the constraint $u^{\mu}u_{\mu}=-1$.
One can investigate the evolution of ESR variables on this spacelike hypersurface with $h_{\mu\nu}$ orthogonal to $u^{\mu}$ i.e. $u^{\mu}h_{\mu\nu}=0$, such that it represents the local rest frame of a freely falling observer in given spacetime (\ref{lineelement}) by using a tensor ${B_{\mu\nu}}$.
For a $n$ dimensional spacetime, ${B_{\mu\nu}}$ can be decomposed as,
\begin{equation}
{B_{\mu\nu}}=\frac{1}{n-1} \, \theta h_{\mu\nu}+\sigma_{\mu\nu}+\omega_{\mu\nu},
\end{equation}
where $\theta$ is the expansion scalar, $\sigma_{\mu\nu}$ represents shear and $\omega_{\mu\nu}$ signifies rotation for geodesic flow.
These variables can be written explicitly as,
\begin{equation}
\theta={B^\mu}_\mu,
\end{equation}
\begin{equation}
\sigma_{\mu\nu}=\frac{1}{2}\left({B_{\mu\nu}}+{B_{\nu\mu}}\right)-\frac{1}{n-1}\theta h_{\mu\nu},
\end{equation}
\begin{equation}
\omega_{\mu\nu}=\frac{1}{2}\left({B_{\mu\nu}}-{B_{\nu\mu}}\right).
\end{equation}
As per their constructional properties, the shear and rotation tensors also satisfy, $h^{\mu\nu}\sigma_{\mu\nu}=0$ and $h^{\mu\nu}\omega_{\mu\nu}=0$ alongwith $g^{\mu\nu}\sigma_{\mu\nu}=0$ and $g^{\mu\nu}\omega_{\mu\nu}=0$.
Since $u^\mu\sigma_{\mu\nu}=0$ and $u^\mu\omega_{\mu\nu}=0$, both $\sigma_{\mu\nu}$ and  $\omega_{\mu\nu}$ are purely spatial in nature (i.e., $\sigma^{\mu\nu}\sigma_{\mu\nu}>0$ and $\omega^{\mu\nu}\omega_{\mu\nu}>0$).
The evolution equation for spatial tensor ${B}_{\mu\nu}$ can also be written as,
\begin{equation}
{\dot{B}_{\mu\nu}}+{B}_{\mu\gamma}{B}^\gamma_\nu=-R_{\eta\nu\mu\delta}u^{\eta}u^{\delta},
\label{eq:Deformation_Tensor_II}
\end{equation}
where $R_{\eta\nu\mu\delta}$ is the Riemann tensor and the dot ($.$) represents differentiation w.r.t. affine parameter $\lambda$.
The trace, symmetric traceless and antisymmetric parts of Eq.(\ref{eq:Deformation_Tensor_II}) leads to the Raychaudhuri equations for ESR variables in four dimensions as below,
\begin{equation}
{\dot{\theta}}+\frac{1}{3}\theta^2+\sigma^2-\omega^2+R_{\mu\nu}u^{\mu}u^{\nu}=0,
\label{eq:REq_I_Expansion}
\end{equation}
\begin{equation}
{\dot{\sigma}_{\mu\nu}}+\frac{2}{3}\theta\sigma_{\mu\nu}+\sigma_{\mu\gamma}\sigma^\gamma_\nu+\omega_{\mu\gamma}\omega^\gamma_\nu+\frac{1}{3}\left(\sigma^2-\omega^2\right)h_{\mu\nu}+\mathcal{C}_{\mu\eta\nu\delta}u^{\eta}u^{\delta}-\frac{1}{2}\tilde{\mathcal{R}}_{\mu\nu}=0,
\label{eq:REq_II}
\end{equation}
\begin{equation}
{\dot{\omega}_{\mu\nu}}+\frac{2}{3}\theta\omega_{\mu\nu}+{{\sigma^\gamma}_\mu}{\omega_{\gamma\nu}}+{{\omega^\gamma}_\mu}{\sigma_{\gamma\nu}}=0.
\label{eq:REq_III}
\end{equation}
where $\sigma^2=\sigma^{\mu\nu}\sigma_{\mu\nu}$, $\omega^2=\omega^{\mu\nu}\omega_{\mu\nu}$, $\mathcal{C}_{\mu\eta\nu\delta}$ is the Weyl tensor and $\tilde{\mathcal{R}}_{\mu\nu}=\left(h_{\mu\gamma}h_{\nu\delta}-\frac{1}{3}h_{\mu\nu}h_{\gamma\delta}\right){R}^{\gamma\delta}$ is the transverse trace-free part of ${R}_{\mu\nu}$.
\subsection{Evolution of ESR variables}
\noindent In order to represent ESR variables at any point in the geodesic congruence associated with timelike vector field $u^{\mu}$, let us consider a freely falling (Fermi) normal frame having the
basis vectors $E_{\eta}^\mu$, $\eta=0,\ldots,3$
(with $E_{0}^\mu = \hat{u}^{\mu}$) which are parallely transported \cite{Dasgupta:2008in,Dasgupta:2012zf}. 
Such frames can be constructed numerically by solving the differential equations $u^\nu\nabla_\nu \, E_{\eta}^{\mu} =0$
(with initial conditions of an orthonormal frame) simultaneously with Eq.(\ref{eq:Deformation_Tensor_II}).
The tensor $B_{\mu \nu}$ in the Fermi basis may then be represented as follows,
$$
B_{\mu \nu} = (\frac{1}{3}\theta +\sigma_{11}) e^1_{\mu}e^1_{\nu} + (\frac{1}{3}\theta +\sigma_{22}) e^2_{\mu}e^2_{\nu} +
(\frac{1}{3}\theta -\sigma_{11}-\sigma_{22}) e^3_{\mu}e^3_{\nu} + (\sigma_{12} - \omega_3)e^1_{\mu} e^2_{\nu} +  \nonumber $$
\begin{equation}
(\sigma_{21} + \omega_3)e^2_{\mu}e^1_{\nu} + (\sigma_{13} + \omega_2)e^1_{\mu}e^{3}_{\nu} + (\sigma_{31} -
\omega_2) e^{3}_{\mu}e^1_{\nu}+ (\sigma_{23} - \omega_1) e^{2}_{\mu}e^{3}_{\nu}+ (\sigma_{32} + \omega_1) e^{3}_{\mu}e^{2}_{\nu}  \, ,\label{bij4d}
\end{equation}
where $e^\eta_\mu$ are co-frame basis which satisfy the relation
$e^\eta_\mu E_\beta^{\mu} = \delta^{\eta}_{\beta}$. The ESR variables
can now be constructed from the evolution tensor (\ref{bij4d}), using the basis vectors $E_\eta^
\mu$, as described in \cite{Dasgupta:2012zf}.
In order to understand the  focusing and defocusing  behaviour of a
timelike geodesic congruence,
let us further redefine the expansion scalar as $\theta = 3{\dot F}/F$.
The Eq. (\ref{eq:REq_III}) may now be expressed in the following Hill-type equation,
\begin{equation}
{\ddot F} + X \, F = 0, \label{theta3e1}
\end{equation}
where $X =(\sigma^2-\omega^2+R_{\mu\nu}u^{\mu}u^{\nu})/3$ with $\sigma^2= 2(\sigma_{11}^2 +
\sigma_{22}^2 + \sigma_{12}^2 + \sigma_{13}^2 + \sigma_{23}^2 + \sigma_{11}
\sigma_{22})$ and $\omega^2 = 2(\omega_1^2 +\omega_2^2 +\omega_3^2)$.
One may note that the Raychaudhuri scalar $R_{\mu\nu}u^{\mu}u^{\nu}$ is zero for the pure SBH case.
It is evident from (\ref{theta3e1}) that for $F\rightarrow 0$ in finite time, we have a finite time singularity in $\theta$ with focusing (defocusing) if $\dot{F}<0$  ($\dot{F}>0$).
The signature of $X$ is thus decisive to examine the focusing/defocusing in view of the critical values for the initial condition on expansion scalar i.e. $\theta_0$ \cite{Dasgupta:2008in,Dasgupta:2012zf}.
When $X$ is positive definite (i.e., $\sigma^2+R_{\mu\nu}u^{\mu}u^{\nu}>\omega^2$),
there exists conjugate points and geodesic
focusing/defocusing takes place accordingly.
On the other hand, no finite time singularity exists for an initially
non-contracting congruence
(i.e., $\theta_0\ge 0$) in case $X$ is negative definite. For $\theta_0<0$, there exists a critical value
below which focusing/defocusing will take place.
\begin{figure}[h!]
\centerline{
\includegraphics[scale=0.28]{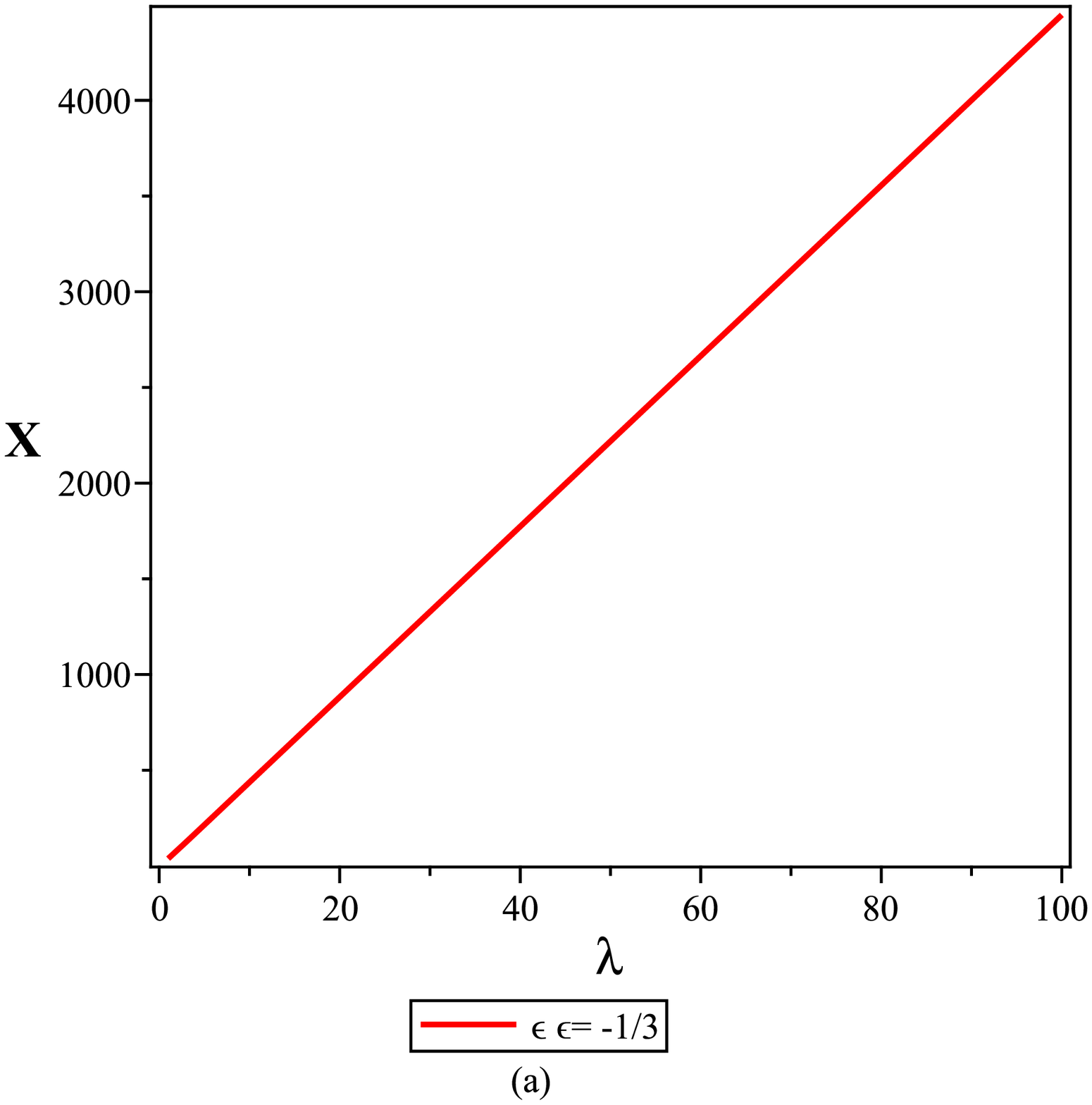}
\includegraphics[scale=0.28]{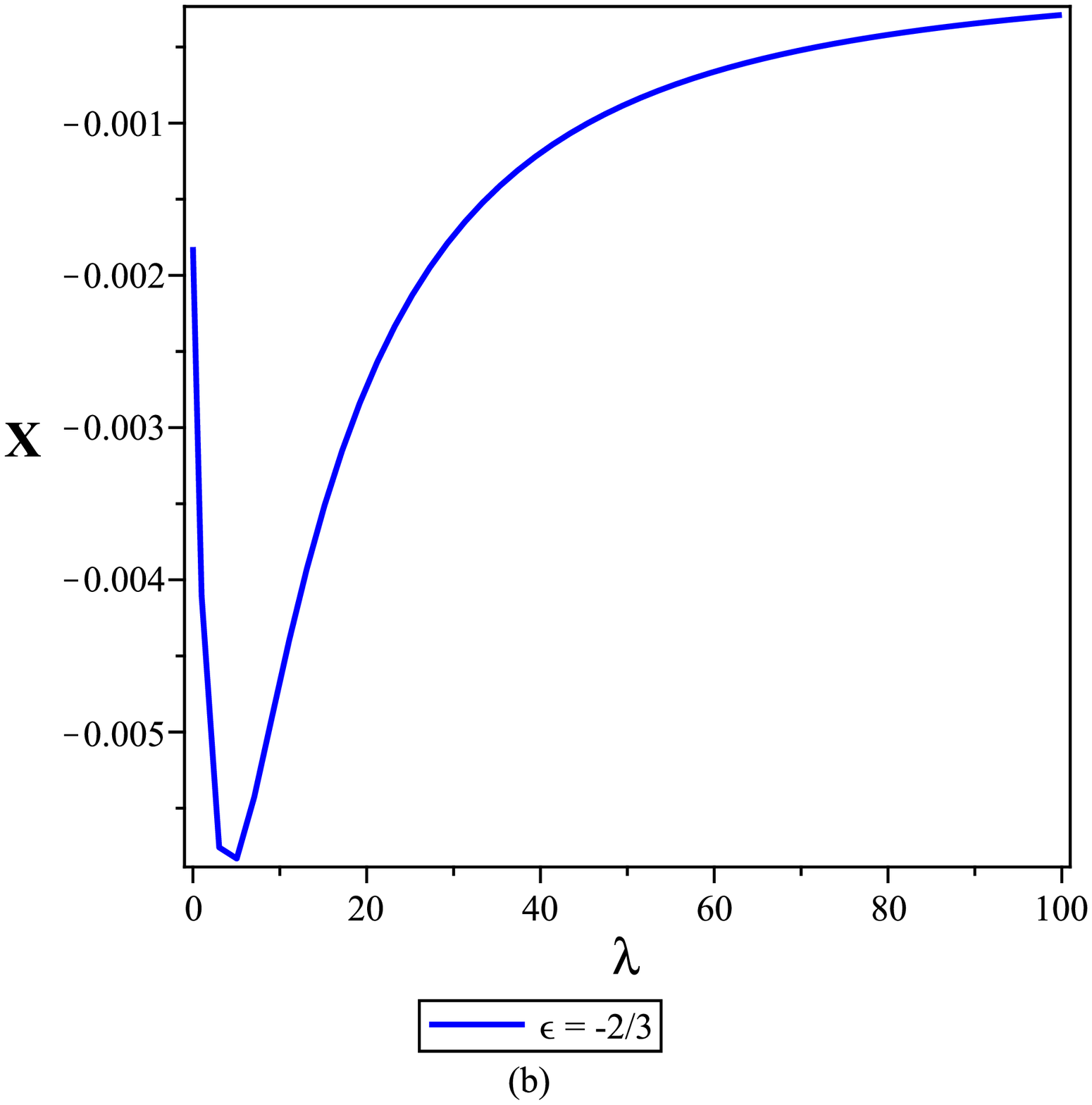}
\includegraphics[scale=0.28]{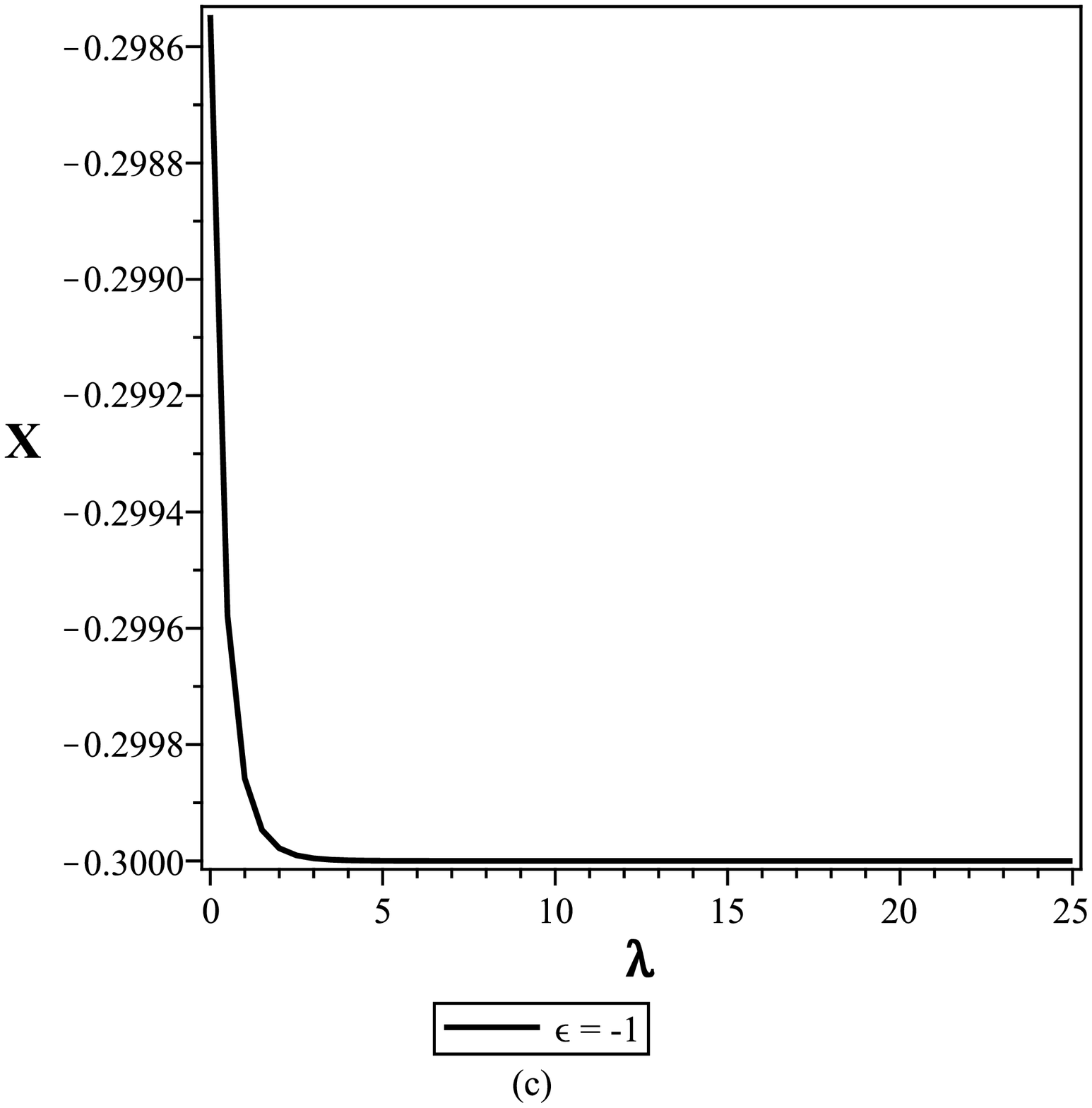}
}
\label{fig:X-focusing}
\caption{Evolution of $X$ with affine parameter $\lambda$, where $E=0.95$, $L=6.5$, $\alpha=0.1$, $M=1$, $Q=0.5$ for different values of $\epsilon$.}
\end{figure}
\noindent From Fig.$3$, one may notice the signature of $X$ easily for different values of $\epsilon$ which in turn reflects the non-attractive behaviour of the quintessential fields in case of $\epsilon=-2/3$ and $\epsilon=-1$ without initial shear and rotation. 
However in view of a positive definite value of $X$ for the case of $\epsilon=-1/3$,  the geodesic focusing may even occur without initial shear and rotation. 
The exact behaviour of geodesic geodesic focusing as well as defocusing can easily be visulaised in the evolution of ESR variables as presented below. 
Using the velocity vector field $u_i$ = ($\dot{t}$,$\dot{r}$,$\dot{\phi}$) in Eqs.(\ref{eq:tg2})-(\ref{eq:velocity_r}), the ESR variables can be represented as functions of $r$.
It leads to the following expression for the expansion scalar,
\begin{equation}
\theta= \pm\,\frac{\left[\left(4r^2(E^2 -1)+6Mr-2Q^2\right)r^{3\epsilon+1}-3\alpha r^2(\epsilon-1)\right]r^{3(\epsilon-1)/2}}{2\left[\left((E^2 -1)r^2+2Mr-Q^2\right)r^{3\epsilon+1}+\alpha r^2\right]^{1/2}}.
\label{eq:theta-r}
\end{equation}
The expression given in Eq.($\ref{eq:theta-r}$) for $\theta$ accommodates the evolution of a geodesic congruence for the fixed value of $E$ and $L$ only. 
It is evident from Eq.($\ref{eq:theta-r}$) that $\theta\rightarrow\pm\infty$ as the denominator vanishes.
It is worth noticing that the denominator of RHS of Eq.(\ref{eq:theta-r}) corresponds to the radial velocity for a test particle with zero-angular momentum and hence it does not include the case of a non-radial motion of test particles as well as the arbitrary initial conditions subjected to the ESR variables.
In order to have a complete analysis of the geodesics deformations, one need to solve the Raychaudhuri Eqs. (\ref{eq:REq_I_Expansion} - \ref{eq:REq_III}) arbitrarily. 
\\
\noindent In the following, the evolution of the ESR variables is presented for the case of a charged BH as well as SBH surrounded with quintessence  background, under the different conditions on the parameters involved. 
The results obtained are compared with those in the RNBH and SBH backgrounds as well as with the corresponding interesting cases having a non-zero cosmological constant.\\
\noindent We study the deformation in equatorial section i.e. $\theta=\pi/2$.
For further numeric computation of the evolution of expansion scalar with affine parameter, we have considered $E=0.95$, $L=6.5$ as it represents the energy and angular momentum per unit mass for the test particle in the innermost circular orbit (ISCO) around a SBH with quintessence when $\epsilon=-2/3$.
For initially diverging congruences (i.e. $\theta_0>0$), there exists a critical value of the expansion scalar ($\theta_c$) below which there is a focusing (i.e. $\theta\rightarrow-\infty$) when the normalisation constant ($\alpha$) and EOS parameter ($\epsilon$) have small magnitude while as the value of $\alpha$ becomes more positive or the value of $\epsilon$ becomes more negative, no focusing is observed as depicted in subsequent figures.
It is important to mention that with the change in the initial conditions on shear and
rotation, the critical value of the initial expansion scalar will also change.
The critical value ($\theta_c$) of the expansion scalar for SBH is calculated as $0.02902$ with $E=0.95$, $L=6.5$ without initial shear and rotation.
For all the figures presented here (Figs. \ref{fig:fig4} - \ref{fig:fig7}), we have considered $\theta_0=0.01$ for $\theta_0<\theta_c$ and $\theta_0=0.1$ for $\theta_0>\theta_c$.
\noindent
Figs.\ref{fig:fig4}(a)-(c) represent the evolution of expansion scalar for $\theta_0<\theta_c$ with $\epsilon$ = -2/3 for different values of $\alpha$, where (a), (b) and (c) represent the case of BH, extremal BH and naked singularity respectively.
Without $\alpha$ geodesics focusing is observed, which converts in defocusing as quintessence appears.
The positive increment in the value of $\alpha$ further supports this defocusing, as geodesic defocusing appears earlier with a positive increase in the value of $\alpha$.
Figs.\ref{fig:fig4}(d)-(f) represent the evolution of expansion scalar for initially diverging geodesics with $\theta_0>\theta_c$   for different values of $\alpha$.
In this case, a positive increment in $\alpha$ first delays the defocusing while if one continuously increases the value of $\alpha$ it then supports the geodesic defocusing.
\newpage
\begin{figure}[h!]
\includegraphics[scale=0.8]{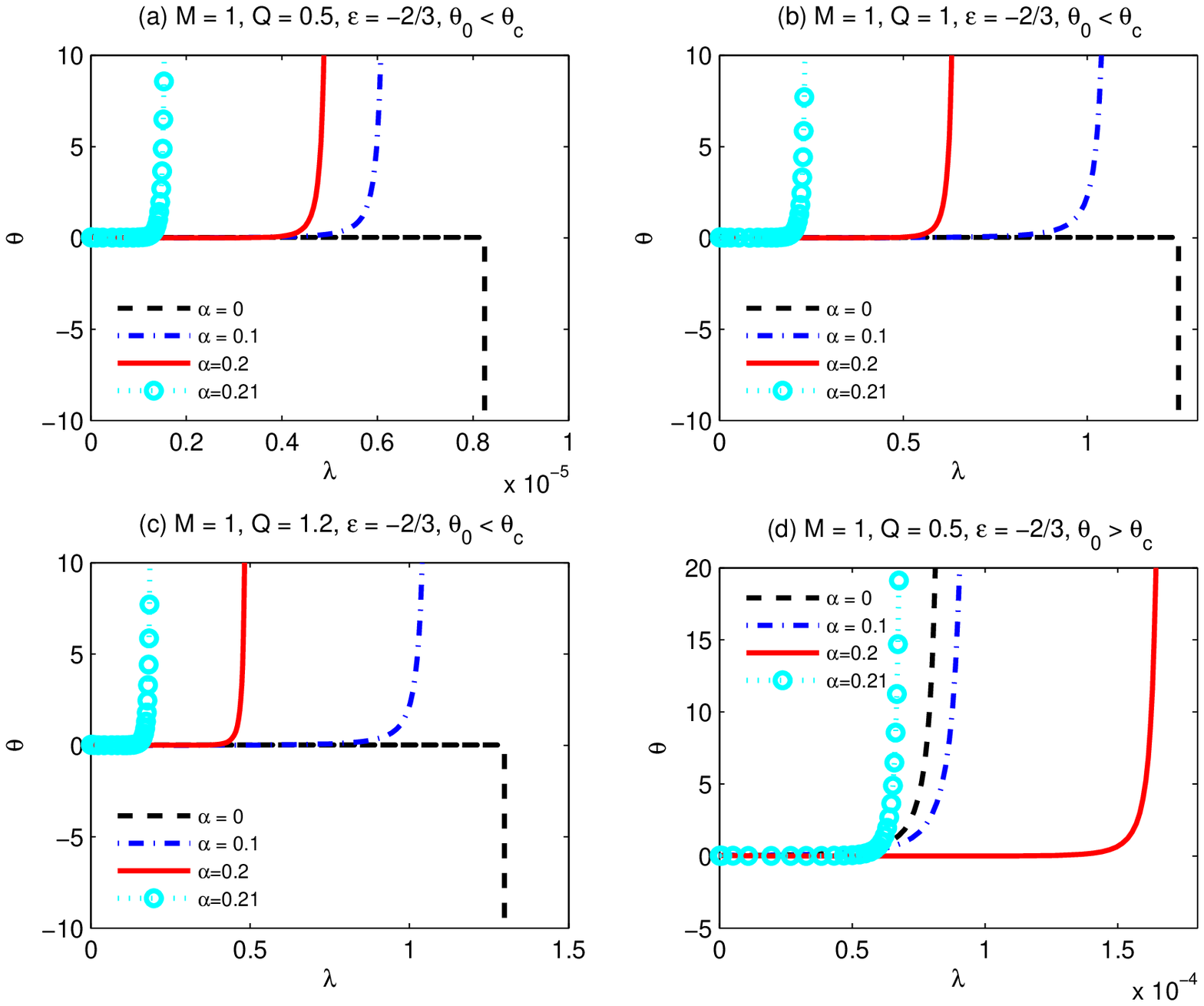}
\nonumber
\end{figure}
\vspace{-2mm}
\begin{figure}[h!]
\includegraphics[width=13.6cm,height=5.4cm]{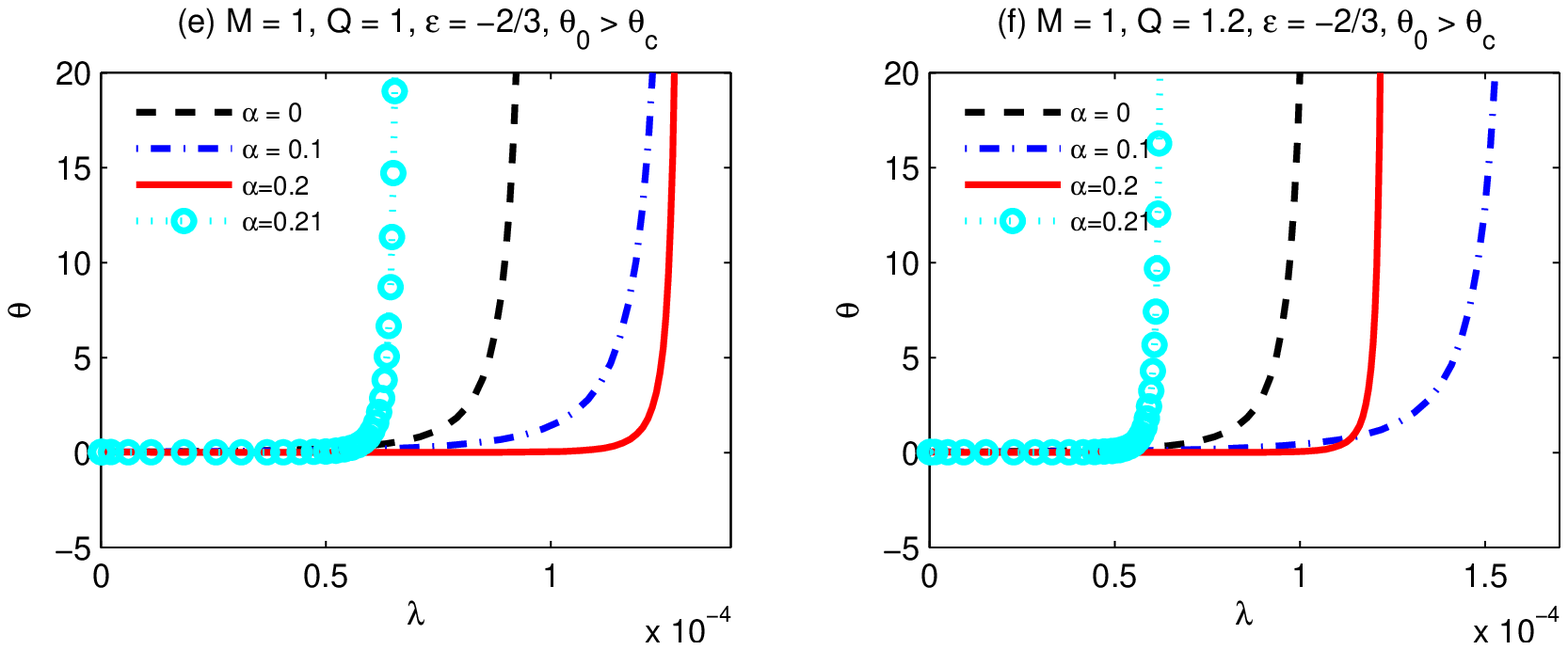}
\caption{The evolution of expansion scalar ($\theta$) with normalisation constant ($\alpha$) with no initial shear and rotation for different values of charge ($Q$).}
\label{fig:fig4}
\end{figure}

\begin{figure}[h!]
\includegraphics[scale=0.9]{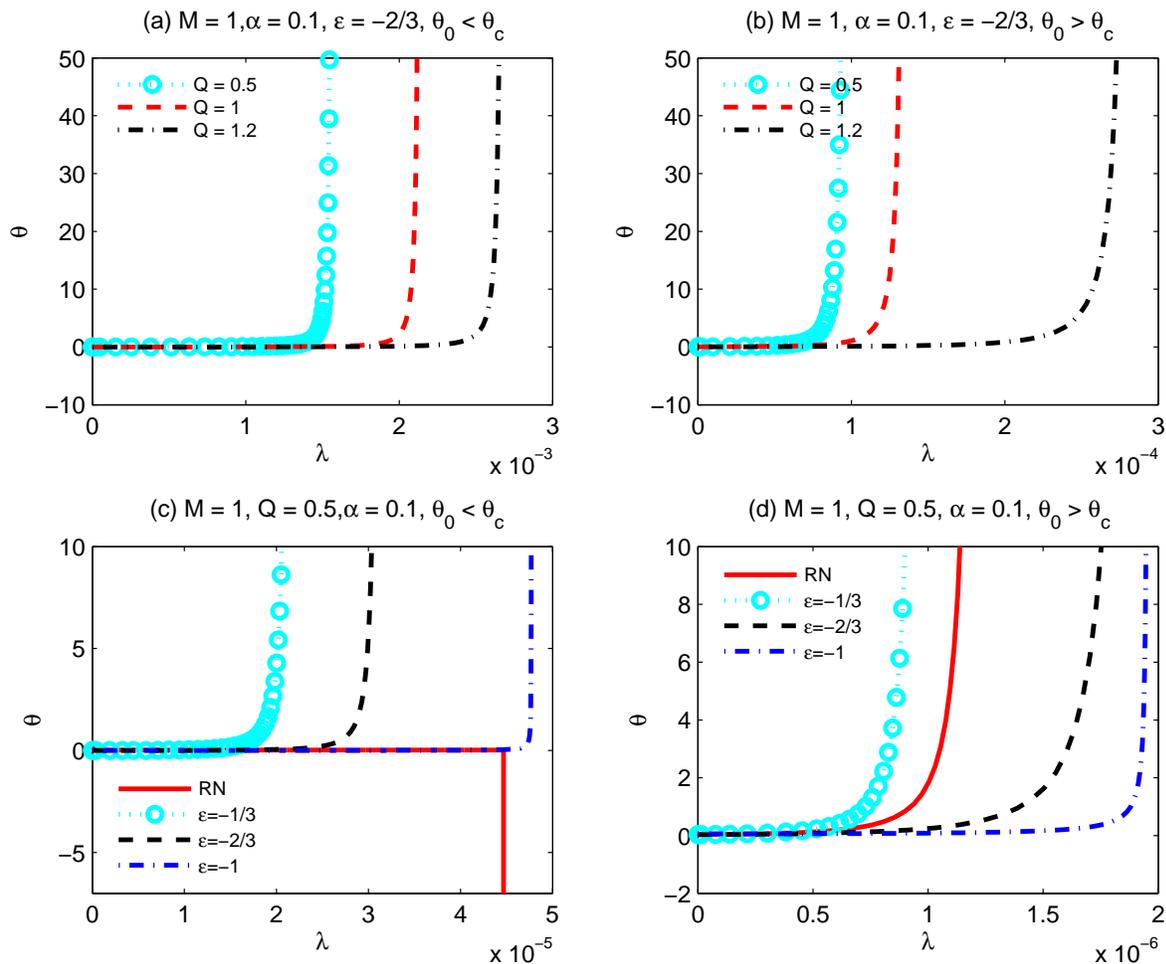}
\caption{The evolution of expansion scalar ($\theta$) with BH charge ($Q$) and EOS parameter ($\epsilon$) with no initial shear and rotation}
\label{fig:fig5}
\end{figure}
\newpage
\noindent Figs.\ref{fig:fig5}(a),(b) depict the effect of increasing BH charge ($Q$) on the evolution of expansion scalar for both the cases of initially diverging geodesics.
It is observed that there occurs no geodesic focusing  due to the presence of quintessence. The defocusing present is further delayed as one compares BH, extremal BH and naked singularity cases, as shown in Fig.\ref{fig:fig5}(a).
Similar is the effect observed with a negative increment in the value of the EOS parameter ($\epsilon$) \cite{Nandan:2009kt} as shown in Figs.\ref{fig:fig5}(c),(d).
\begin{figure}[h!]
\includegraphics[scale=0.8]{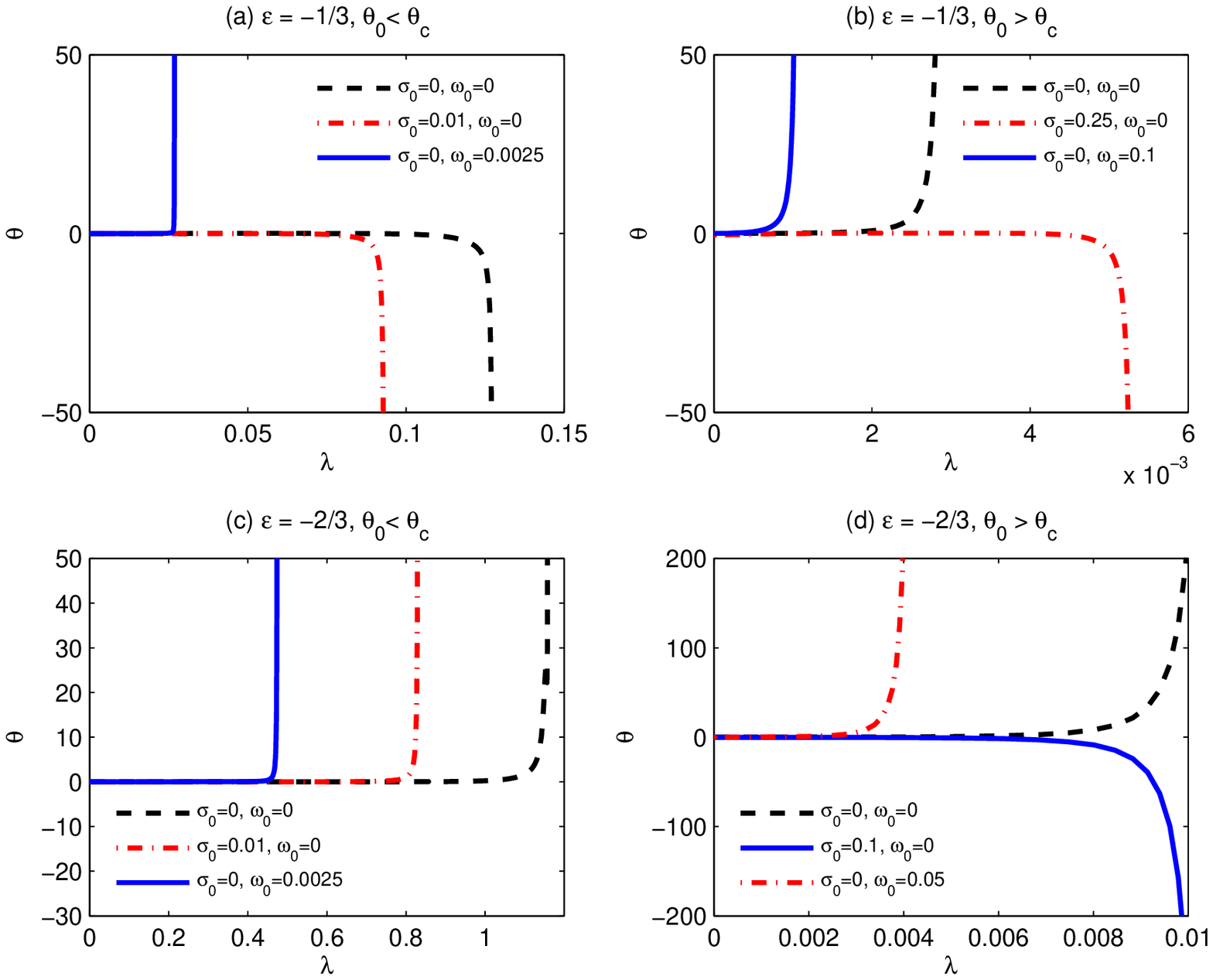}
\includegraphics[width=14.5cm,height=5.5cm]{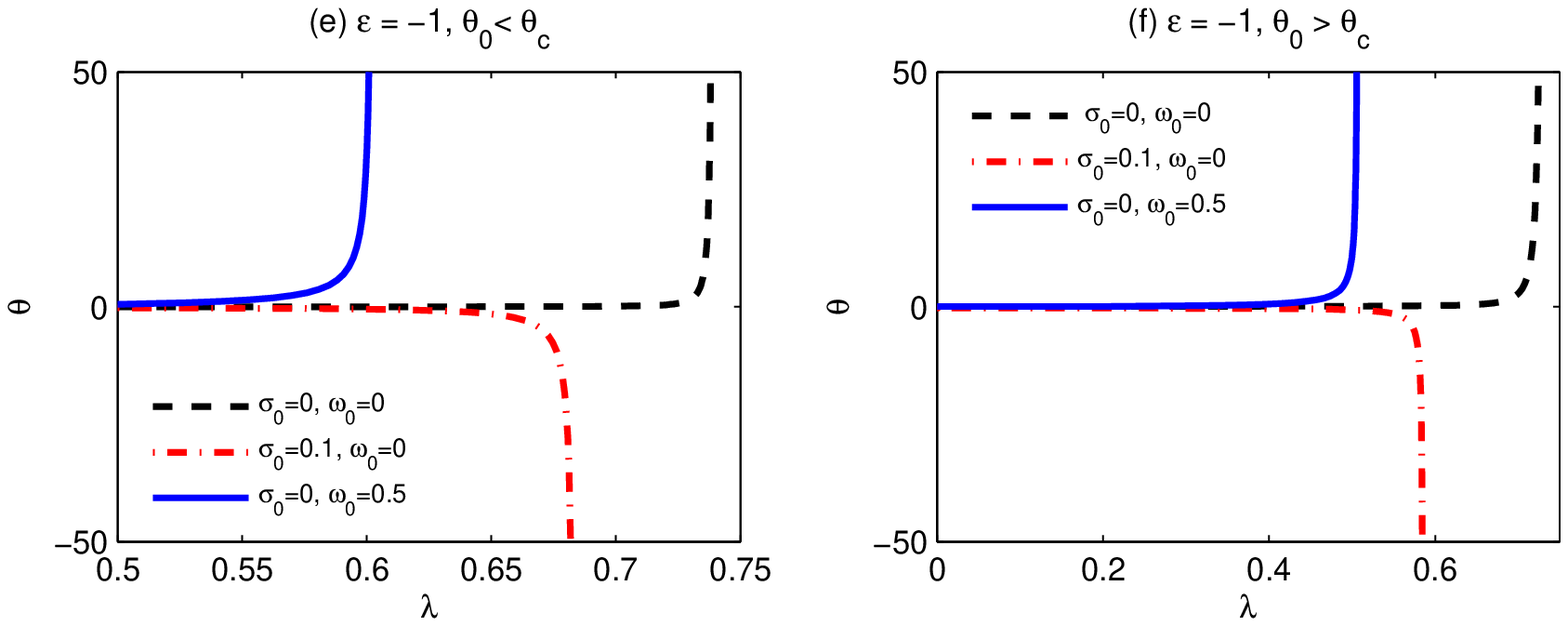}
\caption{The evolution of expansion scalar ($\theta$) with affine parameter ($\lambda$) for different values of EOS parameter ($\epsilon$) and initial conditions on ESR variables.}
\label{fig:fig6}
\end{figure}
\newpage
\noindent Figs.\ref{fig:fig6}(a)-(b) represent the evolution of expansion scalar ($\theta$) for the case $\epsilon$ = -1/3, with fixed value of normalisation constant ($\alpha$).
It clearly shows that without initial shear and rotation geodesics focus (defocus) as $\theta_0$ $<$ $\theta_c$ ($\theta_0$ $>$ $\theta_c$).
Figs.\ref{fig:fig6}(c),(d) and \ref{fig:fig6}(e),(f) represent the evolution of expansion scalar ($\theta$) for $\epsilon$ = -2/3 and -1 respectively, with fixed value of normalisation constant ($\alpha$).
It shows that without initial shear and rotation geodesics defocus even if $\theta_0$ $<$ $\theta_c$.
The presence of an initial shear assists focusing in all the above mentioned cases.
As shown in Fig.\ref{fig:fig6}(a), where focusing is already present, the presence of an initial shear accelerates focusing.
On the other hand, the presence of an initial rotation favours defocusing.
As shown in Fig.\ref{fig:fig6}(b)-(f), in the cases where defocusing is already present, the presence of initial rotation assists it.\\

\begin{figure}[h!]
\includegraphics[scale=0.9]{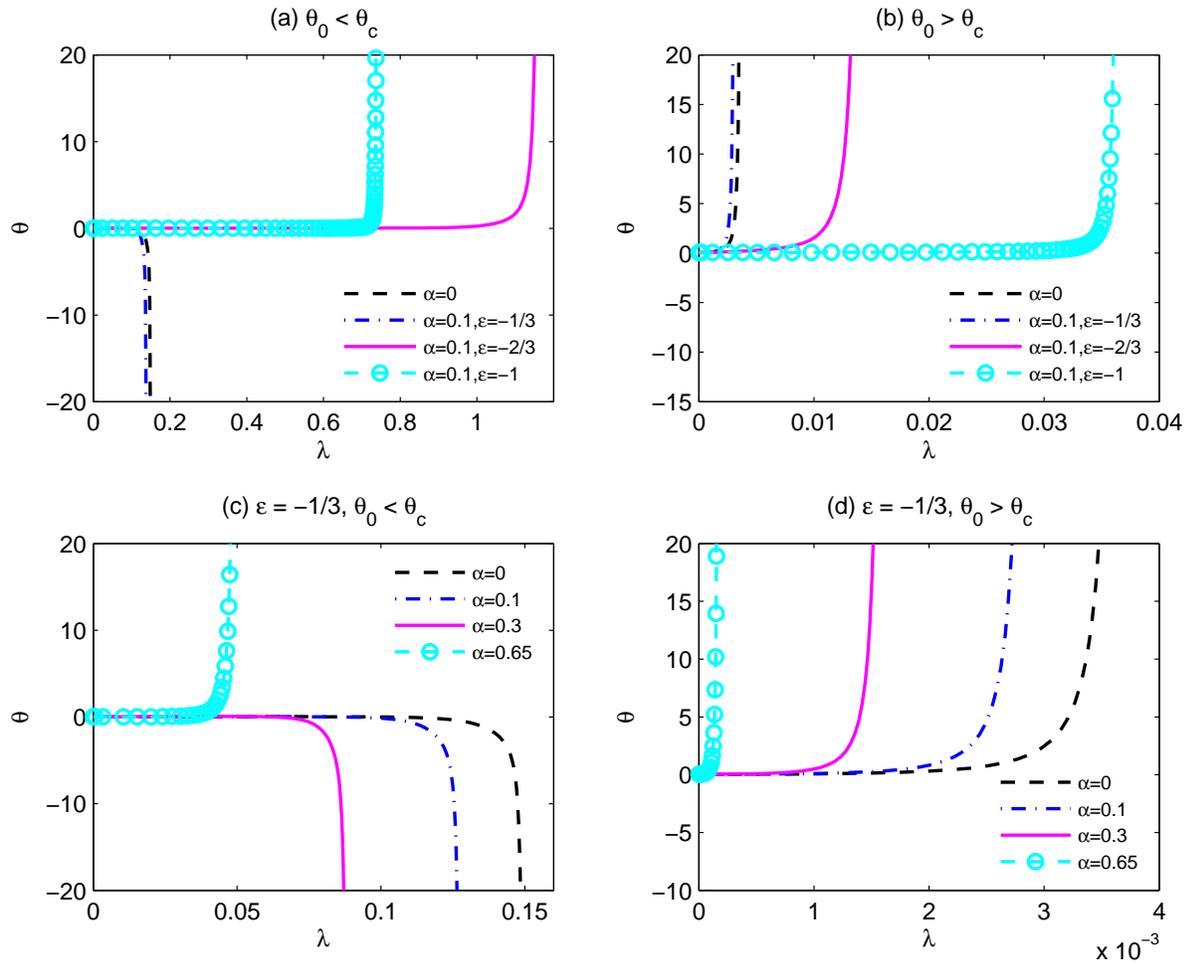}
\caption{The evolution of expansion scalar ($\theta$) with normalisation constant ($\alpha$) without any initial shear and rotation.}
\label{fig:fig7}
\end{figure}
\noindent Figs.\ref{fig:fig7}(a)-(b) represent the comparative plots for the evolution of the expansion scalar ($\theta$) with EOS parameter ($\epsilon$) for SBH surrounded with quintessence.
An increment in the negative value of $\epsilon$ plays different role for both of the cases.
The cases of SBH and $\epsilon$ = -1/3 are quite similar except that the focusing and defocusing both appear earlier for later case due to the non-zero value of $\alpha$.
Figs.\ref{fig:fig7}(c)-(d) represent the effect of normalisation parameter $\alpha$ on the focusing and defocusing of congruences for $\epsilon$ = -1/3.
Infact, an increment in the value of $\alpha$ assists both focusing and defocusing.
\begin{figure}
\includegraphics[width=15.5cm,height=6cm]{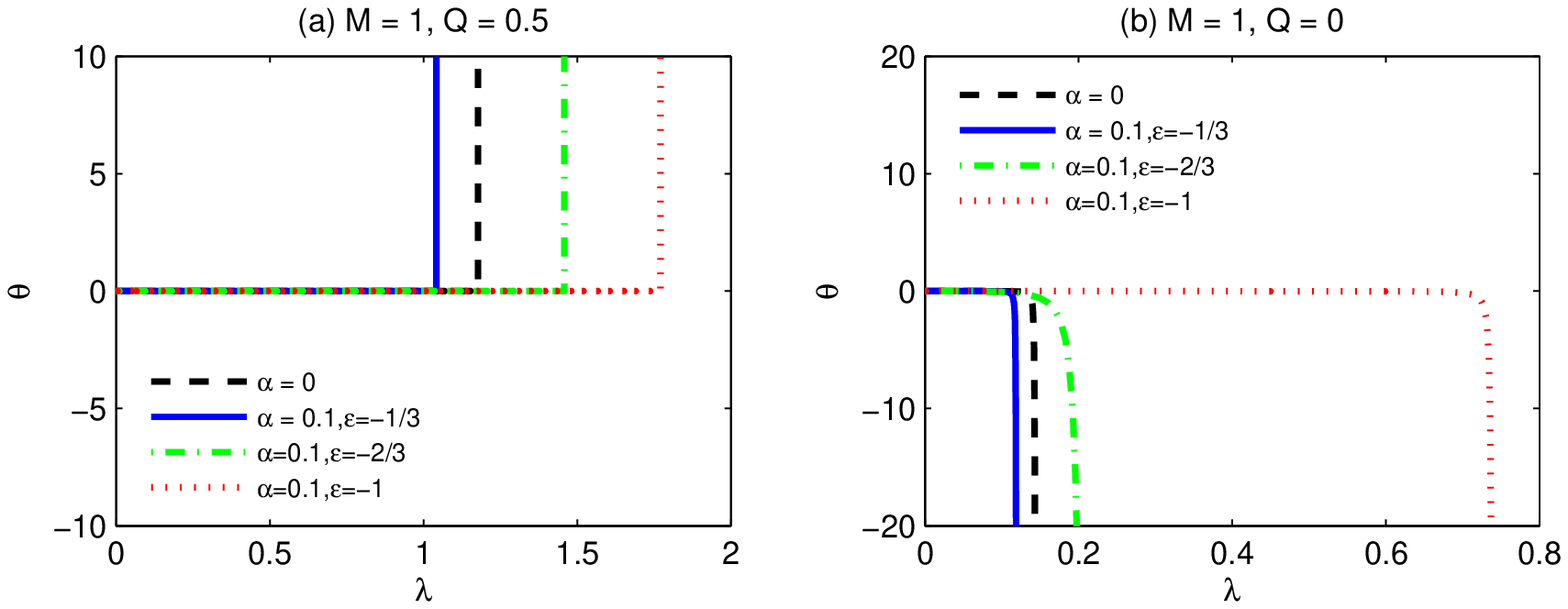}
\includegraphics[width=15.5cm,height=6cm]{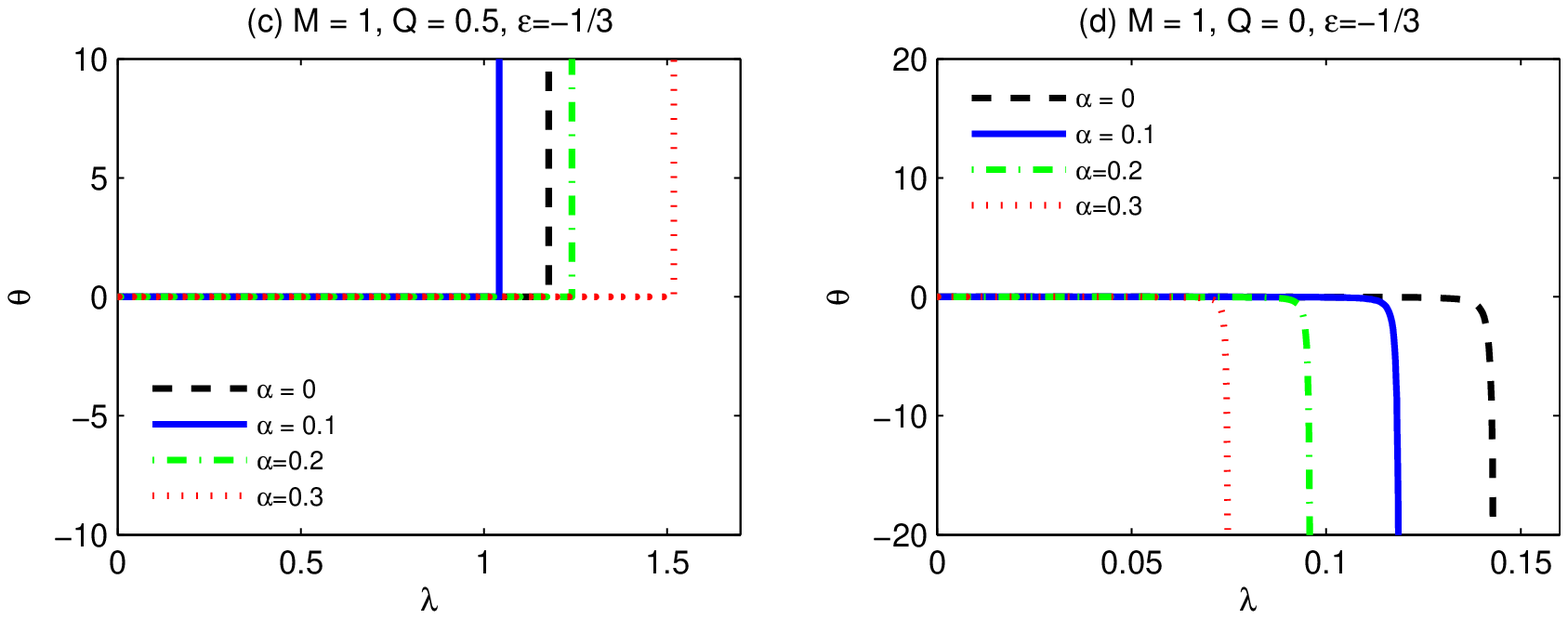}
\caption{The evolution of expansion scalar ($\theta$) with EOS parameter ($\epsilon$) and normalisation constant ($\alpha$) without any initial shear and rotation for initially contracting congruences with $\theta_0=-0.01$.}
\label{fig:fig8}
\end{figure}
The evolution of the expansion scalar ($\theta$) is represented for initially contracting congruences with EOS parameter ($\epsilon$) in Figs.\ref{fig:fig8}(a)-(b) and with normalisation parameter ($\alpha$) in Figs.\ref{fig:fig8}(c)-(d) in the background of RNBH and SBH respectively.
One can notice that the role of negative increment in the $\epsilon$ as well as the positive increment in $\alpha$ is similar to that of the respective cases of initially diverging congruences.
\section{Summary and Conclusions}
\noindent We have investigated the kinematics of of timelike geodesic congruences in the background of a charged BH surrounded with quintessence. The important conclusions are
summarised as follows:
\begin{itemize}
\item The spacetime representing a charged BH surrounded with quintessence satisfies the WEC but voilates the SEC even in the absence of BH charge.
\item The evolution of ESR variables for timelike geodesic congruences is affected qualitatively as well as quantitatively by the normalisation constant and EOS parameter alongwith BH charge and mass.
\item The presence of BH charge supports the defocusing effect of quintessence as for SBH with quintessence the evolution of $\theta$ is similar to the SBH case for small negative values of $\epsilon$ while no focusing is observed in presence of both $Q$ and $\alpha$.
However with the increase in negative $\epsilon$ value, the nature of the evolution shifts towards the corresponding de-sitter spacetimes.
\item The normalisation constant behaves like cosmological constant.
For initially converging congruences, a positive increment in $\alpha$ always assists geodesic focusing.

\item For initially diverging congruences, the positive increment in $\alpha$ assists the defocusing when initial expansion is greater than the its critical value for SBH case. 
However for the other case when the initial expansion is smaller than its critical value, the similar increment in $\alpha$ assists geodesic focusing though the congruences defocus as $\alpha$ value is increased further.\vspace{3mm}\\
The study of the accretion disk formation around the rotating analogue of the BH spacetimes used in this study would be important astrophysically in view of the permissible range of quintessence parameters $\epsilon$ and $\alpha$.
In addition to this, the study of null geodesic flows in the background of such BH spacetimes would be useful to provide a more detailed understanding of these spacetimes.
\end{itemize}

\section*{Acknowledgments}
\noindent The authors are thankful to Prof. M. Sami for useful discussions.
One of the authors HN would like to thank Department of Science and Technology, New Delhi for
financial support through grant no. SR/FTP/PS-31/2009.
\vspace{1cm}

\end{document}